\documentclass[final]{siamltex}
\usepackage{graphicx}
\usepackage{color}
\usepackage{amsmath,amssymb}

\newcommand{\mbh}{\widehat{\boldsymbol{m}}}

\newcommand{\phih}{\widehat{\phi}}

\newcommand{\thetah}{\widehat{\theta}}

\newcommand{\mb}{\boldsymbol{m}}

\newcommand{\ub}{\boldsymbol{u}}
\newcommand{\yb}{\boldsymbol{y}}

\newcommand{\Sigmab}{\boldsymbol{\Sigma}}

\newcommand{\Ub}{\boldsymbol{U}}
\newcommand{\Vb}{\boldsymbol{V}}
\newcommand{\Xb}{\boldsymbol{X}}
\newcommand{\Yb}{\boldsymbol{Y}}

\newcommand{\Fc}{{\cal F}}

\newcommand{\bias}{{\rm Bias}}

\newcommand{\Ex}{{\rm E}}

\newcommand{\fnl}{{f_{\rm N L}}}

\newcommand{\ellp}{{\ell_\phi}}
\newcommand{\mse}{{\rm MSE}}

\newcommand{\oneb}{\boldsymbol{1}}

\newcommand{\R}{{\rm I\!R}}

\newcommand{\Var}{{\rm Var}}

\def\zal2{z_{\alpha/2}}

\newtheorem{result}{Result} 
\begin{document}
\title{Some statistical issues regarding the estimation of $\fnl$ in CMB non-Gaussianity}
\author{L. Tenorio\footnote{Mathematical and Computer Sciences, Colorado School of Mines (ltenorio@mines.edu)} }
\date{\today}
\maketitle

\begin{abstract}
We consider the problem of estimating the parameter $\fnl$ in the standard local model of 
primordial CMB non-Gaussianity. We determine the properties of maximum likelihood (ML)
estimates and show that the problem is not the typical ML estimation problem as there
are subtle issues involved. In particular, The Cramer-Rao inequality is not applicable
and the likelihood function is unbounded with several points of maxima. However, the particular characteristics
of the likelihood function lead to ML estimates that are simple and easy to compute.
We compare their performance to that of moment estimators. We find that ML
is better than the latter for values $\fnl$ away from the origin. For small values of $\fnl$, 
the fifth order moment is better than ML and the other moment estimators.
However, we show how for small $\fnl$, one can easily improve the estimators by a simple
shrinkage procedure. This
is clearly important when the goal is to estimate a very small $\fnl$.   
In the process of studying the inference problem, we address some basic issues regarding
statistical estimation in general that were raised at the Workshop on Non-Gaussianity in Cosmology held in 
Trieste in July 2006. 
\end{abstract}

\pagestyle{myheadings}
\thispagestyle{plain}
\markboth{L.Tenorio}{Statistical estimation of $\fnl$ in CMB non-Gaussianity}

\section{Introduction}
We consider the standard local model of primordial CMB non-Gaussianity where sky maps are modeled as
\begin{equation}\label{eq:local}
Y_i = U_i + \fnl\left(\,U_i^2 - \sigma^2\,\right),
\end{equation}
with $\Ub=(U_1,...,U_n)$ a zero-mean Gaussian vector with covariance matrix $\Sigmab$ and
$\Var(U_i)=\sigma^2$. The covariance matrix is defined by the covariance function
of the random field. The objective is to use the data vector $\Yb=(Y_1,...,Y_n)$ to estimate $\fnl$.

As opposed to the spectrum/moment estimators considered by
Creminelli et al. \cite{crem06} and Babich \cite{babich05}, we will also consider
maximum likelihood and shrinkage versions of the estimators. In addition, we will focus on studying their performance 
under the assumption that the leading order model is correct. That is, while these authors considered
perturbation models of \eqref{eq:local}, we will assume the local model is correct.
The reason is that to understand properties of perturbations, it is important
to understand properties under the ideal case where the leading order model is correct
and standard statistical tools can be used.
Extending the statistical methods to perturbation models is much more difficult;
it is not clear to this author what methods will remain valid and to what degree.  
 
To compare estimators we will mainly use the mean square error so as to take into account bias and variance.
However, to make a proper comparison we will have to address 
some technical issues regarding statistical estimation in general that were raised at the Workshop on 
Non-Gaussianity in Cosmology held in Trieste in July 2006. 

For example, some of the comments and talks at this workshop, as well as in published papers, 
seem to give the message that the Cramer-Rao
(CR) inequality is an ideal way to decide the optimality of estimators; that this inequality can
be used to decide if an estimator provides all the available information about
an unknown parameter, and that one should strive to find unbiased estimators of minimum variance.
However, none of these statements is quite correct. We will show this with simple examples as well
as with the problem of estimating $\fnl$ to test CMB non-Gaussianity. 

In order not to keep repeating the same references for the different statistical
concepts mentioned in this note (in italics), we refer the reader to two excellent general references where all
the definitions can be found: Lehmann \& Casella \cite{lehmann98} and Casella \& Berger \cite{casellaberg}. 

We start by recalling the scalar version of the CR inequality: It states that under `regularity' conditions
the variance of an {\it unbiased estimator} $\thetah(\Yb)$ of $\theta$ satisfies
\[
\Var(\,\thetah(\Yb)\,) \geq \frac{1}{\Fc(\theta)},
\]
where the {\it Fisher information} $\Fc(\theta)$ is defined by
\[
\Fc(\theta)  =  \Var\left(\,\partial/\partial\theta \log L(\theta;\Yb) \,\right)
\]
and $L(\theta;\Yb)$ is the {\it likelihood function}. The first problem when trying to apply this
bound to estimates of $\fnl$ is in the regularity conditions; one of the conditions
is to be able to pass the derivative $\partial/\partial\theta$ of the likelihood
function under the integral
$\int\,dy$. We will see that this cannot be done for the $\fnl$ problem. However, even if the regularity
conditions were satisfied, we would still have to determine if there are unbiased estimators of $\fnl$. The answer again
is negative. Finally, if the conditions were met and there were unbiased
estimators matching the bound, it still would not follow that the estimators would be any good.

But before we consider these problems, we believe it is important to start with two simple examples 
that can be completely worked out and that illustrate some
of the basic issues we will address:\\

\noindent
{\bf Example.} 
Consider the uniform distribution on the interval
$(0,2\theta)$. The goal is to estimate $\theta$. The probability
density function (PDF) is
\[
p(x,\theta) = \frac{1}{2\theta}\,I_{(0,2\theta)}(x),
\]
where $I_A$ stands for the indicator function of the set $A$. The likelihood function of $\theta$ is 
\[
L(\theta;x) = \frac{1}{2\theta}\,I_{(x/2,\infty)}(\theta).
\]
The first two derivatives of the log-likelihood are:
\[
\frac{\partial}{\partial\theta}\,\log L(\theta;x) = \frac{1}{\theta}\,I_{(x/2,\infty)}(\theta),\quad
\frac{\partial^2}{\partial\theta^2}\,\log L(\theta;x) = -\frac{1}{\theta^2}\,I_{(x/2,\infty)}(\theta).
\]
First, note that the derivative $\partial/\partial\theta$ cannot be passed under the integral sign:
\[
\int_{-\infty}^\infty \frac{\partial}{\partial\theta}\,L(\theta;x)\,dx = -\int_0^{2\theta} \frac{1}{2\theta^2}\,dx = 
-\frac{1}{\theta} \neq 0 =\frac{\partial}{\partial\theta} \int_{-\infty}^\infty L(\theta;x)\,dx.
\]
This means that the conditions required for the CR inequality are not satisfied and therefore
it is not necessarily valid. In fact, since the likelihood function is constant in $X$ with probability 1,
it follows that $\Fc(\theta)=0$. Or, formally,
\begin{eqnarray*}
\Fc(\theta) & = & \Var\left(\,\partial/\partial\theta \log L(\theta;X) \,\right)\\
 & = & \Ex\left[\,\left(\,\partial/\partial\theta \log L(\theta;X) \,\right)^2\,\right] -
\left[\,\Ex\left(\,\partial/\partial\theta \log L(\theta;X) \,\right)\,\right]^2\\
&  = & \frac{1}{\theta^2} - \frac{1}{\theta^2} = 0;
\end{eqnarray*}
thus the CR bound is infinite. 
Yet, $\widehat{\theta}(X) = X$ is an unbiased estimator of $\theta$  with finite variance $\theta^2/3$.
The ML estimator is $\widehat{\theta}^{\tiny \rm\, ML}(X) = X/2$. This estimator
is biased but its {\it mean square error} (MSE) is smaller than that of the unbiased estimator:
\begin{eqnarray*}
\mse\left(\,\widehat{\theta}^{\tiny \rm \,ML}(X)\,\right) & = & \Var\left(\,\widehat{\theta}^{\tiny \rm \,ML}(X)\,\right)
+ \bias\left(\,\widehat{\theta}^{\tiny \rm \,ML}(X)\,\right)^2 = \frac{\theta^2}{4}\\
 & \leq &  \frac{\theta^2}{3} = \mse\left(\,\widehat{\theta}(X)\,\right) .
\end{eqnarray*}
More generally, for a sample of size $n$, the sample mean 
$\widehat{\theta}_n(X_1,...,X_n) = \bar{X} = (X_1+\cdots + X_n)/n$
is a {\it consistent estimator} of $\theta$ (i.e., it is unbiased and its variance decreases to 0 as
$n\rightarrow \infty$):
\[
\Ex \left(\,\widehat{\theta}_n(\Xb)\,\right) = \theta,\quad
\Var \left(\,\widehat{\theta}_n(\Xb)\,\right) = \frac{1}{3\,n}\,\theta^2 .
\]

 It is also easy to see that the ML estimator is
$\widehat{\theta}_n^{\tiny \rm \,ML} = X_{(n)}/2$, where $X_{(n)}$ is the largest of the $X_i$. The bias and
variance of this estimator go to zero as the sample size increases
\[
\Ex \left(\,\widehat{\theta}_n^{\rm \,ML} (\Xb)\,\right) =\frac{n}{n+1}\,\theta ,\quad
\Var \left(\,\widehat{\theta}_n^{\rm \,ML}(\Xb)\,\right) = \frac{n}{(n+2)(n+1)^2}\,\theta^2,
\]
which shows that the ML estimate is
also consistent; the requirements for consistency of ML estimates are different from those
of the CR inequality. The variables $X_{(i)}$ are usually called {\it order statistics} and make
their appearance here because the conditions $X_1<2\theta$,...,$X_n<2\theta$ are equivalent to 
the single requirement $X_{(n)}<2\theta$.

The estimator $\widehat{\theta}_n$ is known as a {\it moment estimator} because it is obtained by first solving
for $\theta$ as a function of the moments of $X$ and then substituting the sample moments for the 
population ones (correlation functions, and thus the bispectrum, are examples of moment estimators).
It is well known that moment estimators can usually be improved by conditioning on what are called {\it sufficient
statistics} (functions that contain all the information about $\theta$ that is provided by
the data). Moment estimators are usually not given in terms of sufficient statistics while
ML estimators are. For example, to improve $\widehat{\theta}_n$ we condition on the
sufficient statistic $X_{(n)}$ and obtain the estimator
\[
\widehat{\theta}^c_n\left(\,\Xb\,\right) = \Ex\left(\,\widehat{\theta}_n\left(\,\Xb\,\right)\,|\,X_{(n)}\,\right)=
\frac{n+1}{2 n} X_{(n)}.
\]
This estimator is unbiased and its variance is smaller than that of the original $\widehat{\theta}_n$: 
\[
\Var\left(\,\widehat{\theta}^c_n\left(\,\Xb\,\right)\,\right) = \frac{\theta^2}{4n(n+2)} 
\leq \frac{1}{3n}\,\theta^2 = \Var \left(\,\widehat{\theta}_n\left(\,\Xb\,\right)\,\right)
\]
This confirms that the moment estimate did not contain all the information about $\theta$.

Finally, recall that under regularity conditions $\Fc(\theta)$ can also be computed as
\[
\Fc(\theta) = -\Ex\left(\,\partial^2/\partial\theta^2 \log L(\theta;X) \,\right)
\]
but not for our simple example for the same reason the CR inequality is not valid: we have 
\[
\Fc(\theta) = -\Ex\left(\,\partial^2/\partial\theta^2 \log L(\theta;X) \,\right) = 
1/\theta^2\neq 0.
\]

\noindent
{\bf Example.} We have seen that the CR inequality does not have to hold if the support of the PDF depends on
the parameter. But, could it hold for values of $\theta$ for which the support is `essentially'
$(0,\infty)$? This is not necessarily even in the limit is a finite interval. 
Here is an example: We make a slight change to the PDF 
in the previous example and use the uniform $U(0,\,1 + \theta)$. The support of the PDF still depends on
$\theta$ and so the CR inequality is not valid. But suppose we are
only interested in $\theta \ll 1$. In the limit as $\theta \rightarrow 0$,
the support becomes $(0,1)$. Hence, it may be tempting to conclude
that the CR inequality is `approximately' valid for small $\theta$;
it is not. Just as before, the Fisher information is $\Fc(\theta)=0$ and 
the bound is again infinite. And, just as before, there are unbiased estimators
(e.g., $\thetah_n(\Yb)=2\bar{X}-1$) of finite variance.
The problem of estimating $\fnl$ is even worse because the likelihood function
blows up at the boundaries.\\

The lessons to learn from these simple examples are: (i) The CR bound does not have to be valid
if the support of the PDF depends on the parameters to be estimated; (ii) The CR bound is not always 
applicable and even
when the bound is infinite, there may be consistent unbiased estimators of finite variance; (iii)
an unbiased estimator is not necessarily better than a biased one. In fact, there
are many cases where no unbiased estimators exist, or where a biased estimator is better 
than the best unbiased estimator.
Furthermore, although not shown in this example,
there are many examples where an unbiased estimator has a variance that matches the CR
bound and yet its MSE is larger than that of a biased estimator; (iv)
moment estimators (such as the bispectrum) do not usually contain all the information
about the unknown parameter that the data provide and therefore can be
improved by conditioning on sufficient statistics. 

We will show that the problem of estimating $\fnl$ is similar to that
of estimating $\theta$ in the examples above but
it is slightly worse in that the likelihood function is unbounded, it has more than one maximizer
and there are no finite variance unbiased estimators of $\fnl$.

The rest of the note is organized as follows: 
In Section \ref{sec:toy}, we start with a particular case of the local model \eqref{eq:local} 
to show that the conditions for the 
CR inequality of $\fnl$ are not satisfied. We determine some of the particular characteristics
of the likelihood function and derive simple expressions for estimates based on maximum likelihood.
In Section \ref{sec:gen} we return to the general local model \eqref{eq:local} 
to extend the results found for the simple model. Again we find simple expressions for
the ML estimates but find that neither the moment nor the ML estimators are better
for all values of $\phi$. We then make shrinkage modifications to the estimators
to improve their performance for small values of $\fnl$. Some technical details 
are left to the Appendix.

\section{The toy case}
\label{sec:toy}
We now return to the local model \eqref{eq:local} and start with a particular case 
described in Section III of Babich \cite{babich05}. This model is based on independent and identically distributed
observations
\begin{equation}\label{eq:Y}
Y_i = U_i + \fnl (U_i^2 - \sigma^2),
\end{equation}
where the $U_i$ are independent $N(0,\sigma^2)$.
For simplicity, we will write $\phi$ instead of $\fnl$.

\subsection{PDF and likelihood function}\label{sec:pdf}

To derive the PDF and study the CR bound, it is enough to consider the case of a single 
$Y=U + \phi (U^2 - \sigma^2)$, which can be written as
\[
Y =\phi\sigma^2\,V - \left(\,\frac{1}{4\phi} + \phi\sigma^2\,\right),
\]
where 
\[
V = \left(\,\frac{U}{\sigma} + \frac{1}{2\phi\sigma}\,\right)^2.
\]
By definition $V$ has a {\it noncentral} $\chi_1^2$ distribution with noncentrality
parameter $\lambda = 1/(4\sigma^2\phi^2)$ and thus its PDF is
\[
p(V,\lambda) = \frac{1}{\sqrt{2\pi}}\,\frac{e^{-(v+\lambda)/2}}{\sqrt{v}}\,\cosh\left(\,\sqrt{\lambda v}\,\right)\,I_{v\geq 0}.
\]
The PDF of $Y$ can be easily derived as $Y$ is just a rescaled and shifted $V$. 
The problem is the condition $V\geq 0$; 
this boundary makes the support of the PDF
of $Y$ depend on the unknown parameter $\phi$. The PDF can be written as
\[
p(y,\phi) = p^-(y,\phi)(y)\,I_{\phi<0} + p^+(y,\phi)(y)\,I_{\phi \geq 0},
\]
where 
\[
p^\pm(y,\phi)(y) = \frac{|\,\phi\,|^{-1/2} }{\sqrt{2\pi\sigma^2}}\,
\frac{e^{-(y+\phi\sigma^2 + 1/2\phi)/(2\phi\sigma^2)}}{\sqrt{|\,y+\phi\sigma^2 + 1/4\phi\,|}}
\,\cosh \left(\,\frac{\sqrt{|\,y+\phi\sigma^2 + 1/4\phi\,|}}{2\sigma^2\,|\,\phi\,|^{3/2}}\,\right)
\,I_{A^\pm}
\]
with the sets $A^\pm$ defined as $A^+ = \{\,y > -\phi\,\sigma^2 - 1/4\phi\,\}$ and 
$A^- = \{\,y < -\phi\,\sigma^2 - 1/4\phi\,\}$.

In particular, for $\phi>0$ the support in $y$ of $p(y,\phi)$ has a left boundary defined by $A^+$, which depends on
$\phi$. This is a problem for the CR inequality because the derivative $\partial/\partial\phi$ 
cannot be passed under the integral $\int dy$. In fact, having a PDF whose support does
not depend on the unknown parameters is often stated as a requirement for the CR
inequality (e.g., Bickel \& Doksum \cite{bickeldoksum}). This is an important point that
cannot be overlooked.

In our toy problem the support of $p(y,\phi)$ not only depends on $\phi$ but the likelihood function also
diverges to infinity at the boundaries of $A^\pm$. However, one can still define reasonably
good ML-based estimates.
In fact, we will show below that the boundaries actually
lead to simple estimates that in the general case depend only on $\sigma^2$ and not
on the full covariance matrix. 
\begin{figure} 
\centerline{\includegraphics[width=9cm,angle=0,keepaspectratio=1]
{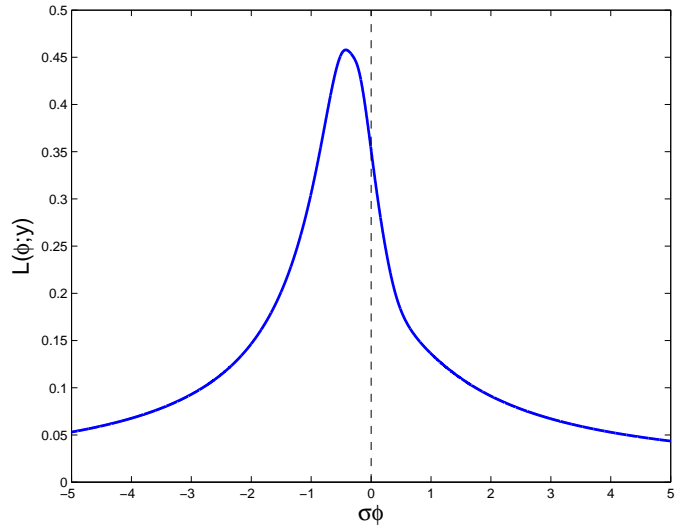}}
\caption{Likelihood function \eqref{eq:maxL} for $y=1/2$.
\label{fig:likfun}}
\end{figure}
\subsection{Maximum likelihood estimates}
The shape of the likelihood function depends on the value of $y$. On the region $\phi>0$ 
the likelihood is not zero only if $y$ is such that $\phi$ belongs to $A^+$; on $\phi<0$
we need a y such that $\phi$ belongs to $A^-$. It is easy to see that when
$|\,y\,| < \sigma$ (or when $|\,y\,|<1$ if we normalize the data to estimate
$\phi\sigma$), the likelihood function is just 
\begin{equation}\label{eq:maxL}
L(\phi,y) = p^+ (y,\phi)\,I_{\phi>0} + p^- (y,\phi)\,I_{\phi<0}.
\end{equation}
This function has a unique maximum and the ML estimate is well defined. For example, Figure \ref{fig:likfun}
show the likelihood function of $\phi\sigma$ for $y=1/2$.

The case $|\,y\,|\geq \sigma$ is more pathological. For $\phi>0$ and $y\leq -\sigma$, we need
\[
\phi < \frac{|\,y\,| - \sqrt{y^2 - \sigma^2}}{2\sigma^2}\quad \mbox{or}\quad
\phi > \frac{|\,y\,| + \sqrt{y^2 - \sigma^2}}{2\sigma^2},
\]
which delimits the boundary of $A^+$. Since the likelihood function blows up at the two limits, it
is maximized by setting $\phi$ to be either of the limits. Hence the maximum likelihood
estimate is not well defined. A similar thing happens for $\phi<0$ and $y\geq \sigma$.
However, this does not mean that neither of the two possibilities
is good. In fact, we get a reasonably good estimate by averaging them:
Define the ML-based estimate of $\phi$ as
\[
\phih^{\rm\, ML}_1(\,Y\,) = \left\{ \begin{array}{ll}
|\,Y\,|/2\sigma^2 & \mbox{if $Y\leq -\sigma$}\\
\mbox{maximizer of \eqref{eq:maxL}} & \mbox{if $|\,Y\,|<\sigma$}\\
-|\,Y\,|/2\sigma^2 & \mbox{if $Y\geq \sigma$}.
\end{array}\right. 
\]
It may seem strange that the estimate is positive when $Y$ is negative and conversely (it also
happens in Figure \ref{fig:likfun}) but it can
be explained heuristically as follows: If $Y<-\sigma$, ten $U + \phi(U^2-\sigma^2)<-\sigma$, that is,
\begin{equation}\label{eq:sign}
\phi\,(U+\sigma)(U-\sigma) < -(\sigma +U).
\end{equation}
But since $U$ is Gaussian $N(0,\sigma^2)$, it is 
in the interval $-\sigma<U<\sigma$ with probability 68\%. Hence, about 70\% of the time
we expect $\sigma+U > 0$ and $U-\sigma <0$ and thus the right hand side in \eqref{eq:sign}
is negative and for the left side to have the same sign we need $\phi>0$. Similarly, we
need $\phi<0$ when $Y>\sigma$.

A better estimate is obtained by taking in each case the quadratic root closest to zero
\[
\phih^{\rm \,ML}_2(\,Y\,) = \left\{ \begin{array}{ll}
\left(\,|\,Y\,| - \left(\,Y^2 - \sigma^2\,\right)^{1/2}\,\right)/2\sigma^2 & \mbox{\quad if $Y\leq -\sigma$}\\
\mbox{maximizer of \eqref{eq:maxL}} & \mbox{\quad if $|\,Y\,|<\sigma$}\\
\left(\,-|\,Y\,| + \left(\,Y^2 - \sigma^2\,\right)^{1/2}\,\right)/2\sigma^2 & \mbox{\quad if $Y\geq \sigma$}.
\end{array}\right. 
\]
In the section below we compare the MSE of these two estimators to that of some moment estimates.

\subsection{Moment estimates}
We now compare ML to moment estimators of $\phi$. To make our point, it is again sufficient to consider the 
scalar case of a single $Y$. 

We know that the odd moments of a Gaussian variable $U\sim N(0,\sigma^2)$ are zero while the even moments
are given by
\[
\Ex \,U^{2m} = \frac{ (2m)! }{2^m m!}\,\sigma^{2m} = C_{2m}\,\sigma^{2m}.
\]
It follows that
\begin{eqnarray}
\frac{1}{\sigma^{2m}}\,\Ex\,Y^{2m} & = & \sum_{2k\leq 2m} E_{2m,2k}\,\left(\,\phi\sigma\,\right)^{2k}\label{eq:moments2m}\\
\frac{1}{\sigma^{2m+1}}\,\Ex\,Y^{2m+1} & = & \sum_{2k+1\leq 2m+1} E_{2m+1,2k+1}\,\left(\,\phi\sigma\,\right)^{2k+1}
\label{eq:moments2mp1},
\end{eqnarray}
where the sums are over even and odd indices, respectively, and
\[
E_{m,k} = \begin{pmatrix} m \\ k \end{pmatrix}
\sum_{j=0}^k \begin{pmatrix} k \\ j \end{pmatrix} (-1)^{k-j} \,C_{m-k+2j} . 
\]
As we have assumed that $\sigma$ is known, we focus on estimators of the dimensionless
parameter $\phi\sigma$ using the normalized random variable $Y/\sigma$.
For example, the first two estimators based on the third and fifth moments are:
\begin{equation}\label{eq:momest}
\phih^{(3)}\left(\,Y/\sigma\,\right) = \frac{1}{E_{3,1}}\,(Y/\sigma)^3,\qquad \phih^{(5)}\left(\,Y/\sigma\,\right) = 
\frac{1}{E_{5,1}}\,(Y/\sigma)^5.
\end{equation}
The question we want to address now is whether $\phih^{(3)}$ is better than any of the
other moment estimators. We use the MSE to compare $\phih^{(3)}$, $\phih^{(5)}$
and the ML estimator.  
The variance and bias of $\phih^{(3)}$ and $\phih^{(5)}$ are easily 
derived using \eqref{eq:moments2m} and \eqref{eq:moments2mp1}. The MSE of the ML estimates
is calculated through simulations.

\begin{figure} 
\centerline{\includegraphics[width=9cm,angle=0,keepaspectratio=1]
{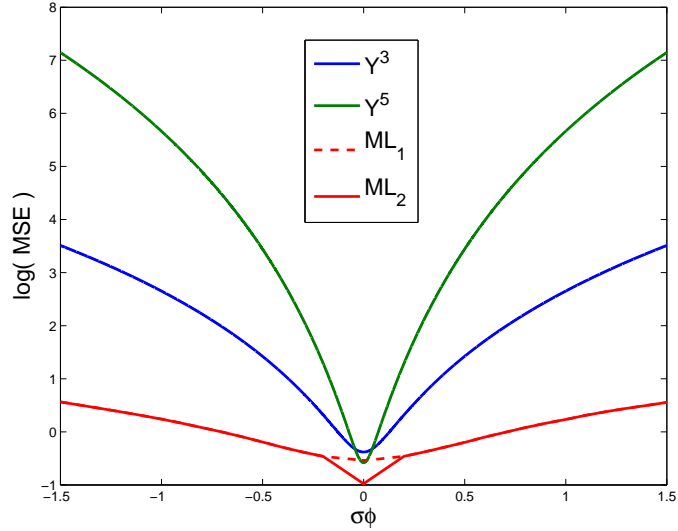}}
\caption{Mean square error (MSE) of the moment and ML estimators of $\sigma\phi$ for the
toy model \eqref{eq:Y}.
\label{fig:momest}}
\end{figure}

Figure \ref{fig:momest} shows the MSE of $\phih^{(3)}$, $\phih^{(3)}$ and the two ML estimators
$\phih^{\rm \,ML}_1$ and $\phih^{\rm \,ML}_2$. We see that $\phih^{(3)}$ is better than $\phih^{(5)}$
except for small values $|\,\sigma\phi\,|<0.05$. For values $|\,\sigma\phi\,|<0.02$, $\phih^{(5)}$
is better than the first ML based estimate $\phih^{\rm\, ML}_1$. The ML estimate $\phih^{\rm \,ML}_2$
has much smaller MSE than the moment estimates. However, we will see that the general case is not
as clear cut.

\subsection{Non-existence of unbiased estimators of $\phi$}

In the simple example discussed in the introduction, the CR bound is infinite but there are finite
variance unbiased estimators of the unknown parameter. However,
unbiased estimators do not always exist and our toy model provides one such example. 

Suppose there is a finite variance unbiased estimator $\phih(\Yb)$ of $\phi$ based on $\Yb$.
That is,
\[
\Ex\, (\,\phih(\Yb)\,) = \phi,\qquad \Var (\,\phih(\Yb)\,) < \infty
\]
for all $\phi \in \R$ and any $\sigma>0$. Then, by dividing each $Y_i$ in
\eqref{eq:Y} by a constant $\alpha\neq 0$ we obtain 
\[
\tilde{Y}_i = \tilde{U}_i + \tilde{\phi}\,\left(\,\tilde{U}_i^2 - \tilde{\sigma}^2\,\right),
\]
where $\tilde{Y}_i = Y_i/\alpha$, $\tilde{\sigma} = \sigma/|\,\alpha\,|$ and
$\tilde{U}_i = U_i/\sigma \sim N(0,\tilde{\sigma}^2)$. But since $\phih(\Yb)$ is an unbiased estimator,
we must have $\Ex\left(\,\phih(\tilde{\Yb})\,\right)=\tilde{\phi}$ and therefore 
\begin{equation}\label{eq:phial}
\phih_\alpha(\Yb) = \frac{1}{\alpha}\,\phih\left(\,\frac{\Yb}{\alpha}\,\right)
\end{equation}
is an unbiased estimator of $\phi$ for any $\alpha\neq 0$. This implies that 
\[
\int \phih(\yb) p(\alpha y_1,\phi)\cdots p(\alpha y_n,\phi) \, d\yb = \frac{\phi}{\alpha^{n-1}}
\]
for any $\alpha\neq 0$.
Hence, the inverse of the left hand side has to be a monomial in $\alpha$ of degree $n-1$.
In particular, it is a constant for $n=1$. We consider the case $n=1$ in the appendix and show
that the condition is impossible to meet. Hence, there are no unbiased estimators of
finite variance based on $Y$. Of course, one would still have to check if there are
unbiased estimators based on more that one $Y_i$ (the number of such $Y_i$ is called the
{\it degree} of the estimator). 

\section{The general case}\label{sec:gen}
We now return to the general case \eqref{eq:local}.
As in the scalar case, a simple transformation leads to
\[
\Yb = \phi\,\Vb - \oneb\,\left(\,\frac{1}{4\phi} + \sigma^2\,\phi\,\right),
\]  
with
\[
\Vb = \left(\,\Ub + \frac{\oneb}{2\phi}\,\right).*\left(\,\Ub + \frac{\oneb}{2\phi}\,\right)
\]
where `.*' stands for point-wise multiplication, and $\oneb=(1,...,1)$. The PDF of $\Vb$ is of the form
\[
p(\Vb,\phi) = F(\Vb,\phi)\,I_{V_1\geq 0}\cdots I_{V_n\geq 0},
\]
for some `nice' exponential function $F(\Vb,\phi)$ that will be defined below. It follows that the
PDF of $\Yb$ is
\begin{equation}\label{eq:pdfy1}
p(\Yb,\phi,\Sigmab ) = \frac{1}{|\,\phi\,|^n}\,F\left(\,\frac{\Yb + \ell_\phi\,\oneb}{\phi},\phi\,\right)
I\left(\,\frac{Y_1 + \ell_\phi}{\phi}\geq 0\,\right)\cdots I\left(\,\frac{Y_n + \ell_\phi}{\phi}\geq 0\,\right), 
\end{equation}
where $\ell_\phi = 1/4\phi + \sigma^2\phi$. 

Just as in the scalar case, the support of the PDF of $\Yb$ depends on $\phi$; it is defined
by the two sets
\[
B^+ = \left\{\,Y_{(1)} \geq -\ell_\phi\,\right\},\qquad B^- = \left\{\,Y_{(n)} \leq -\ell_\phi\,\right\},
\] 
where $Y_{(1)}$ and $Y_{(n)}$ denote the smallest and largest values of the sample. Hence,  
the CR bound is again not applicable but just as in the toy local model, there are simple
expressions for estimates of $\phi$ based on maximum likelihood. 

\begin{figure} 
\centerline{\includegraphics[width=14cm,angle=0,keepaspectratio=1]
{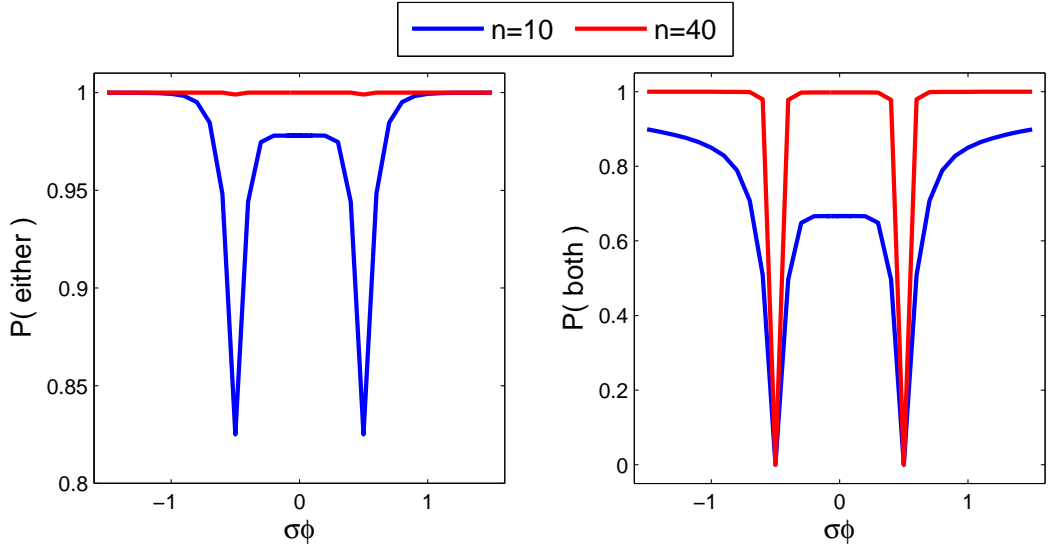}}
\caption{The left panel shows the probability that either $Y_{(1)}\leq -1$ or $Y_{(n)}\geq 1$
as a function of $\sigma\phi$ for independent samples of sizes $n=10$ and $n=40$ of the local model \eqref{eq:local}.
The right panel shows the probability that $Y_{(1)}\leq -1$ and $Y_{(n)}\geq 1$.
\label{fig:orderprobs}}
\end{figure}
\subsection{Maximum likelihood estimates}
To find ML estimates of $\phi$ in the general case, we first need to derive the function $F(\Vb,\phi)$
mentioned in the previous section.
This is easy to do by computing the probability $P\,[\,V_1 \leq v_1,...,V_n \leq v_n\,]$. The final result is:
\[
F(\Vb,\phi) \propto \frac{1}{\sqrt{v_1}\cdots\sqrt{v_n}}\, G(\Vb,\phi,\Sigmab ),
\]
\[
G(\Vb,\phi,\Sigmab ) = \sum_{k_1,...,k_n=0}^1 g\left(\,\ell_\phi + (-1)^{k_1}\sqrt{v_1},...,\ell_\phi + 
(-1)^{k_1}\sqrt{v_1},\Sigmab\,\right)
\]
\[
g(\ub,\Sigmab ) = \frac{1}{|\,\Sigmab\,|^{n/2}}\,e^{-\ub^t\Sigmab^{-1}\ub/2}.
\]
To be consistent with Section \ref{sec:pdf}, write the likelihood function of $\phi$ as
\[
L(\phi;\Yb,\Sigmab) = p^{-}(\Yb,\phi,\Sigmab) I_{\phi<0} +  p^{+}(\Yb,\phi,\Sigmab) I_{\phi> 0}
\]
with
\[
p^{\pm}(\Yb,\phi,\Sigmab) \propto \frac{1}{|\,\phi\,|^n} \frac{1}{\Pi_i\sqrt{y_i+\ell_\phi/\phi}}\,
G\left(\,\frac{\Yb + \ell_\phi\,\oneb}{\phi},\phi,\Sigmab\,\right)\,I_{B^\pm} .
\] 
As in the scalar case, to find the maximum likelihood estimates we need to consider different cases.
For $Y_{(1)}\leq -\sigma$ the likelihood blows up at the roots of $\sigma^2\phi^2
+ \phi Y_{(1)} + 1/4=0$, and there are two values of $\phi$ for which this happens. A similar thing
happens with $Y_{(n)}\geq \sigma$ and the two roots of  $\sigma^2\phi^2 + \phi Y_{(n)} + 1/4=0$.
What is good about these cases is that they do not require the use of the full covariance
matrix $\Sigmab$, only the variance $\sigma^2$ is needed. 
The covariance matrix is needed only
when $|\,Y_{(1)}\,|$ and $|\,Y_{(n)}\,|$ are both less than $\sigma$. In this case the likelihood 
remains finite and has to be maximized.  

However, in a large sample one expects $Y_{(1)}$ to
be small and $Y_{(n)}$ to be large. Thus the case $|\,Y_{(1)}\,|, |\,Y_{(n)}\,|\leq \sigma$ 
should be rare in large samples. On the other hand, depending on $\phi$, it is not always possible
to have $Y\leq -\sigma$. For example, it cannot happen for $\phi=1/2\sigma$. However, with high
probability $Y_{(1)}\leq -\sigma$ or $Y_{(n)}\geq \sigma$ and in either case the
ML estimates are roots of the quadratic equations that only require $\sigma^2$. 

For example, the left panel
in Figure \ref{fig:orderprobs} shows the probability that $Y_{(1)}\leq -\sigma$ or $Y_{(n)}\geq \sigma$
based on independent samples of size $n=10$ and $n=40$. We see that with only $n=10$, the probability
is at least 82\% that either $Y_{(1)}\leq -\sigma$ or $Y_{(n)}\geq \sigma$ . With $n=40$
the probability is very close to one. The right panel shows
the probability that both cases will happen. We see that for large samples the two cases
do not happen together only when $\sigma\phi\approx \pm 1/2$. 
But of course, in the general case the samples $Y_i$ are correlated because $U$ has a full covariance matrix
but for sky maps with thousands of pixels we expect more than 40 `degrees of freedom' so we
still expect the same behavior of small and large values of $Y_{(1)}$ and $Y_{(n)}$, respectively.

\begin{figure} 
\centerline{\includegraphics[width=15cm,angle=0,keepaspectratio=1]
{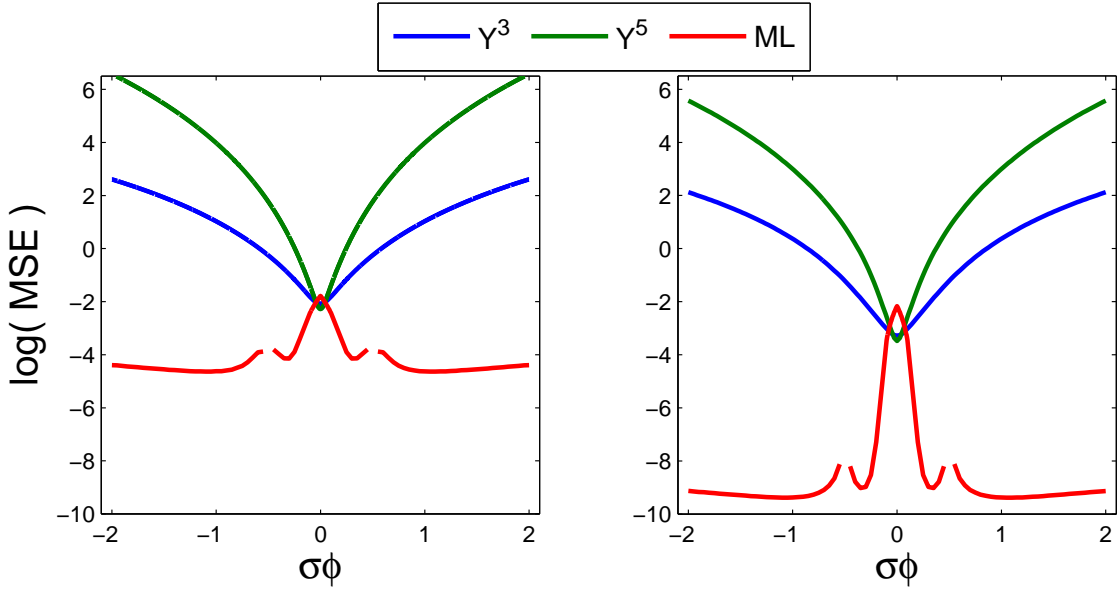}}
\caption{MSE of $\phih^{(3)}$, $\phih^{(5)}$ and the ML estimator \eqref{eq:ML}. The left panel
shows the results for a sample size $n=50$ and the right panel for $n=800$.
\label{fig:momestgen}}
\end{figure}

The selection of the appropriate quadratic roots depends on the range of values of $\sigma \phi$ under consideration.
However, a simple geometric argument can be used to select them: if $\phi>0$, then, as a function of $U$, the parabola
$U+\phi(U^2-\sigma^2)$ has a 
minimum at $U=-1/2\phi$ with minimum $Y$ value  
\[
Y_{\rm min}= -\frac{1}{4\phi}-\sigma^2\phi 
\]
and therefore $\phi$ is a solution of $\sigma^2\,\phi^2 + \phi\,Y_{\rm min} + 1/4=0$. There are two solutions
of this equation, depending on whether $\phi>1/2\sigma$ or $0<\phi<1/2\sigma$. These solutions
are functions of $Y_{\rm min}$.  We have a similar
situation when $\phi<0$, in which case we find $\phi$ as a function of $Y_{\rm max}$. Hence, we have the following
ML estimator
\begin{equation}\label{eq:ML}
\phih^{\rm \,ML}(\,\Yb\,) = \left\{ \begin{array}{ll}
r^{-}_{-} & 0\leq \phi <1/2\sigma\\
r^{-}_+ & \phi>1/2\sigma\\
\pm 1/2\sigma & \phi = \pm 1/2\sigma\\
r^{+}_+ & -1/2\sigma\leq \phi < 0\\
r^{+}_{-} & \phi<-1/2,
\end{array}\right. 
\end{equation}
where the roots are defined as
\[
r^{-}_\pm = \frac{\,|\,Y_{(1)}\,| \pm \left(\,Y_{(1)}^2 - \sigma^2\,\right)^{1/2}}{2\sigma^2},\quad
r^{+}_\pm = \frac{\,-|\,Y_{(n)}\,| \pm \left(\,Y_{(n)}^2 - \sigma^2\,\right)^{1/2}}{2\sigma^2} .
\]
For example, let us return to the case where the $Y_i$ are independent. Define the third and fifth moment
estimates as in \eqref{eq:momest} but averaging over the $n$ observations.
Figure \ref{fig:momestgen} shows the MSE of the ML and moment estimates for $n=50$ (left panel)
and $n=800$ (right panel). We make the following observations:
(i) The fifth and third moment estimates, especially the former,
provide better estimates for values $|\,\sigma\phi\,|<0.05$ but otherwise the ML estimator
has much smaller MSE; (ii) The decrease in the MSE with increasing sample size is much better for the ML estimators.
This decrease is hardly noticeable in the moment estimates for larger values of $\sigma\phi$ where
the estimates are dominated by the bias. 

\begin{figure} 
\centerline{\includegraphics[width=15cm,angle=0,keepaspectratio=1]
{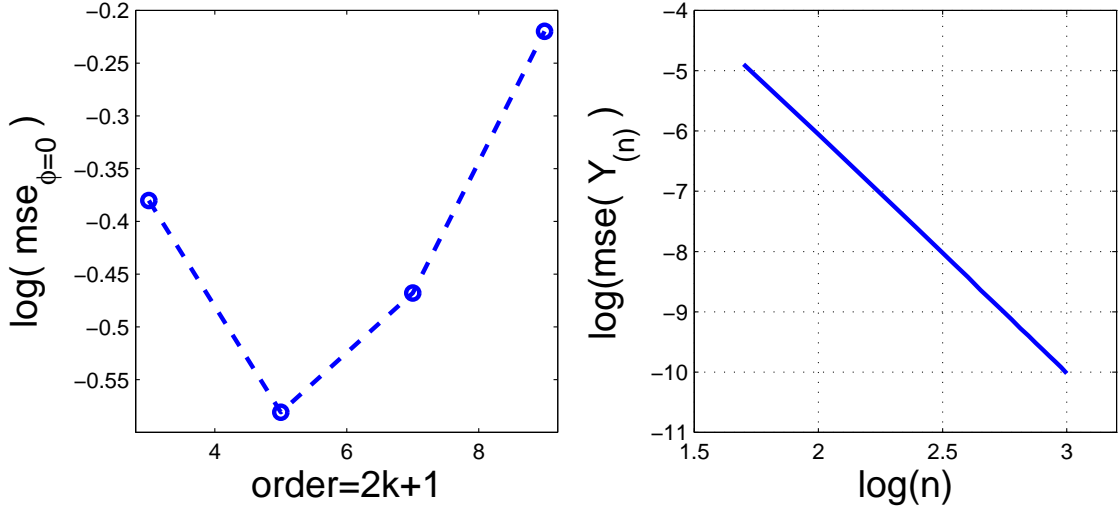}}
\caption{The left panel shows the limit, as $\phi\rightarrow 0$, of the MSE of moment estimators $\phih^{(2k+1)}$ as
a function of order $2k+1$. It shows that $\phih^{(5)}$ has the smallest MSE near $\phi=0$.
The right panel shows the decrease with sample size $n$ of the MSE of $Y_{(n)}$ as an estimator of $Y_{\rm max}$ 
(as defined in the text). It explains the decrease of the MSE of the ML estimates in Figure
\ref{fig:momestgen}.
\label{fig:mommse}}
\end{figure}

Near $\phi=0$ the smallest MSE is that of $\phih^{(5)}$ because as $\phi\rightarrow 0$, the MSE
of $\phih^{(2k+1)}$ is 
\[
{\rm MSE}(\,\phih^{(2k+1)}\,)_{\phi=0}= \frac{E_{4k+2,0}}{n\, E_{2k+1,1}^2}.
\]
The left panel in Figure \ref{fig:mommse} shows the plot of this MSE (for fixed $n$) as a 
function of moment order $2k+1$;
the minimum value is at $2k+1=5$. Thus, for small $\phi$, $\phih^{(5)}$ is better than $\phih^{(3)}$ and the other
higher order moment estimators.

The large MSE of $\phih^{\rm \, ML}$ near the origin comes from ML providing estimates that
are, on the average, greater than $\sigma\phi$ (in absolute value).
From the Appendix, we know that there are problems near the origin. It is also clear
that as $\phi\rightarrow 0$ the parabola, used to choose the roots, becomes a line and the problem is ill-posed. 
It should be possible to improve ML estimates near the origin using {\it shrinkage estimators} or
{\it penalized likelihood}. We return to this below.

The fast decrease in MSE of the ML estimate is interesting and can be explained
as follows. Estimating $\phi$ reduces to estimating $Y_{\rm min}$ (for $\phi>0$) and $Y_{\rm max}$
(for $\phi<0$) with $Y_{(1)}$ and $Y_{(n)}$, respectively. The convergence of the corresponding
estimates is then explained by the convergence of $Y_{(1)}$ to $Y_{\rm min}$ and
$Y_{(n)}$ to $Y_{\rm max}$, respectively. For example, the right panel in Figure \ref{fig:mommse} shows the MSE of 
$Y_{(n)}$ as an estimator of $Y_{\rm max}$ as a function of $n$ for $\sigma\phi=-1$. We see 
the same decrease (as $1/n^4$) in MSE as in Figure \ref{fig:momestgen}.
It is not difficult to study the convergence rate and asymptotic distributions of 
$Y_{(n)}$ and $Y_{(1)}$ using extreme value theory (e.g., Galambos \cite{galambos}).\\

\subsubsection{Performance for small $\phi$}
\begin{figure} 
\centerline{\includegraphics[width=15cm,angle=0,keepaspectratio=1]
{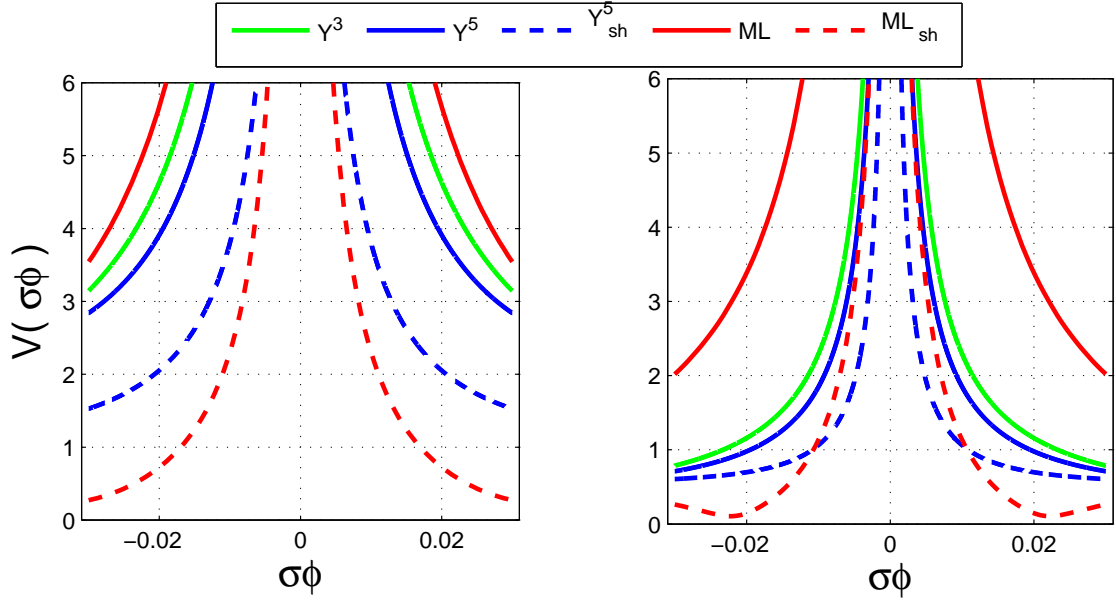}}
\caption{Same as Figure \ref{fig:momestgen} but using the relative measure of
error $V(\phi) = {\rm MSE}^{1/2}/|\,\phi\,|$ (nstead of MSE)
for $n=50$ and $n=800$. The dashed lines show `shrunk ' versions of $\phih^{(5)}$ and
$\phih^{\rm\, ML}$ as defined in the text.
\label{fig:momestgenT}}
\end{figure}

We now focus on small values of $\phi$ to determine if any of the estimators discussed above is better than
the simple zero estimator: $\phih_0(\Yb)=0$ . Figure \ref{fig:momestgenT} 
shows the results for $|\,\sigma\phi\,|<0.03$ and sample sizes $n=50$ (left) and
$n=800$ (right). This time we show a relative measure
of the error defined by $V(\phi) = {\rm MSE}^{1/2}/|\,\phi\,|$ ($\phi\neq 0$). 
Estimators that are better than $\phih_0$ should
have $V<1$. The figure shows that with $n=50$ neither of the three estimators improves on the zero estimator.
For the larger sample, $\phih^{(5)}$ and $\phih^{(3)}$ are an 
improvement but slight and only for $|\,\sigma\phi\,|>0.02$. However, this is not the best we can do.
Our estimators can be improved near the origin using ideas based on shrinkage estimators.

For example, to estimate
a mean vector $\mb$ with observations $\Yb = \mb + \mbox{noise}$, one can minimize
$\|\,\Yb -\mbh\,\|^2 + \lambda\,\|\,\mbh\,\|^2$ (where $\lambda>0$ is fixed) over $\mbh$ and end up with 
$\mbh = \Yb/(1+\lambda)$. This estimator `shrinks' toward the origin and it is the result
of minimizing a goodness of fit norm that `penalizes' vectors of large norms. This is a typical procedure used
to obtain solution of ill-posed inverse problems (e.g., O'Sullivan \cite{sullivan86}, 
Tenorio \cite{tensiam}). Another example is the well-known {\it James-Stein shrinkage estimator} of a 
multivariate Gaussian mean, which improves on
the minimum variance unbiased estimator.

The idea is then to define estimators of the form $\phih_{\rm sh}(\Yb)=(1-\lambda(\Yb,n))\,\phih(\Yb)$, where 
$\lambda(\Yb,n)$ takes values in the interval $(0,1)$. We get the estimators $\phih$ and $\phih_0$, respectively, 
in the limits as
$\lambda\rightarrow 0$ and $\lambda\rightarrow 1$.
The problem is then to select a good shrinkage factor. In our case, even the simple choice of
a constant factor $\lambda$ already leads to improved estimates. We have chosen the factor
that gives the best results for each estimator. The dashed lines in Figure
\ref{fig:momestgenT} show the value of $V$ for shrinkage versions of $\phih^{(5)}$ and
$\phih^{\rm, \, ML}$. For the small sample, only the shrunk ML improves on the zero estimator
and only for $|\,\sigma\phi\,|>0.017$. For the larger sample, the best results are those of the shrunk
$\phih^{(5)}$ and $\phih^{\rm \, ML}$. Both improve on $\phih_0$ for $|\,\sigma\phi\,|>0.011$ but
ML leads to a much smaller MSE. One should be able to get even better results with more sophisticated
choices of $\lambda(\Yb,n)$. 

\section{Summary}
We have compared higher order moments to maximum likelihood (ML) estimators of $\fnl$ and found that no estimator
is better than the others over all values of $\phi$. The selection of an appropriate estimator will
depend on the relevant range of values of $\phi$. For very small values, the fifth moment leads
to the estimator with smallest mean square error (MSE), followed by the third moment and ML.
But we have also shown that even for small $\fnl$, the fifth moment is not the best
as all the estimators considered here improve with the introduction of a shrinkage factor.
In practical applications, the difference between the estimators will depend on the size of the sample and the
covariance structure that determines its `effective degrees of freedom'. The selection of shrinkage
factors will also depend on the particular covariance structure and sample size.

We found that away from the origin the ML estimate has the smallest MSE. 
However, while we have found a very fast decay of
the MSE with increasing sample size, the asymptotics considered here were for the case of independent observations. 
A correct asymptotic analysis should study the in-field asymptotics; an excellent
example with applications to cosmology is Marinucci \cite{marinucci06}.
For finite samples, the performance of the estimators should be studied for the particular 
covariance matrices of the cosmological random fields under consideration. In addition, a more careful
study should determine the robustness of the estimators against deviations from the local
model coming from higher order terms.

We have compared estimators mainly through the MSE and its decrease with increasing sample
size. We do not think that the Cramer-Rao (CR) inequality is the appropriate tool
to select optimal estimators: matching the CR bound is not enough to guarantee a good estimator.
For example, shrinkage estimators are not found by matching a CR bound. 

We are aware of a certain bias in favor of unbiased estimators in the community.
However, all the estimators we have considered here are biased. I hope this note helps us
remember that:
Unbiased estimators do not have to exist. When they exist, they may be difficult
to find. If we find one, it may not be very good.\\

\noindent
{\bf Acknowledgments}. The author thanks the organizers of the Workshop on Non-Gaussianity
in Cosmology (Trieste, July 2006) for an interesting and well organized meeting. He is also
grateful to D. Babich, P.A. Martin, W. Navidi and P.B. Stark for helpful comments.

\bibliographystyle{siam}

\appendix
\section{Proof of the non-existence of unbiased estimators}

We assume that $\phih$ is a finite variance unbiased estimator of $\phi$ based on $Y$ and that
$\phih_\alpha$, as defined in \eqref{eq:phial}, is an unbiased estimator of $\phi$
for any $\alpha>0$ and reach a contradiction. This will show that there are
no finite variance unbiased estimators of $\phi$ based on $Y$. 

It is enough to consider $\phi>0$. To simplify notation, we define the following functions
\[
\ellp = \phi\sigma^2 + \frac{1}{4\phi}
\]
\[
G(\alpha,y) = e^{-(\alpha y+\ellp + 1/4\phi)/(2\phi\sigma^2)}\,
\cosh\left(\,\frac{1}{2\sigma^2\,\phi^{3/2}}\,\sqrt{\,\alpha y+\ellp\,}\,\right)
\]

\[
H(\alpha,y)=
\frac{G(\alpha,y)}{\sqrt{\,\alpha y+\ellp\,}}.
\]
With this notation
\begin{equation}\label{eq:exphial}
\Ex\, \phih_\alpha(Y) = \sqrt{\frac{1}{2\pi\phi\sigma^2}}\,\int_{-\ellp/\alpha}^\infty H(\alpha,y)\,\phih(y)\,dy.
\end{equation}

\begin{result}{\rm For any $y> 0$, $H(\alpha,y)$ is a decreasing function of $\alpha>0$ and
\[
\lim_{\alpha\rightarrow \infty} H(\alpha,y) = 0.
\]
}
\end{result}
\noindent
{\em Proof:} Define 
\[
h(\alpha,y) = \alpha y + \ellp = \alpha y +\phi\sigma^2 + \frac{1}{4\phi} \geq \frac{1}{4\phi}. 
\]
The  we can write
\[
H(\alpha,y) = \frac{1}{\sqrt{h(\alpha,y)}}\,e^{-(h(\alpha,y) + 1/4\phi)/(2\sigma^2\phi)}
\cosh\left(\,\frac{\sqrt{h(\alpha,y)}}{2\sigma^2\phi^{3/2}}\,\right).
\]
Clearly, the first factor decreases as $\alpha$ increases, as does the second factor when multiplied
by the decreasing exponential of the hyperbolic cosine. All we have left to show is that
\[
H(\alpha,y) = e^{-(h(\alpha,y) + 1/4\phi)/(2\sigma^2\phi)}
e^{\sqrt{h(\alpha,y)}/(2\sigma^2\phi^{3/2})}.
\]
is also decreasing. To show this, it is enough to show that the following difference decreases
\begin{eqnarray*}
D(\alpha,y) & = & \frac{1}{\phi^{3/2}}\,\sqrt{h(\alpha,y)} - \frac{1}{\phi}\,(\,h(\alpha,y) + 1/4\phi\\
 & = & \frac{1}{\phi^2}\,\left(\,\sqrt{\phi\,h(\alpha,y)} - (\,\phi\,h(\alpha,y) + 1/4\,)\,\right) < 0 .
\end{eqnarray*}
But since $\phi\,h(\alpha,y)\geq 1/4$, we have
\[
\frac{\partial}{\partial\alpha}\, D(\alpha,y) = \frac{\phi y}{2\sqrt{\phi\,h(\alpha,y)}}
\,\left(\,1 - 2\sqrt{\phi\,h(\alpha,y)}\,\right) < 0 .
\]
Hence $D(\alpha,y)$ is also a decreasing function of $\alpha$ and so is $H(\alpha,y)$.
Finally, since the first exponential in the definition of $H(\alpha,y)$ has a negative exponent
and $D(\alpha,y)<0$ as shown above, it follows that
\[
0\leq H(\alpha,y) \leq \frac{2}{\sqrt{h(\alpha,y)}} \rightarrow 0
\] 
as $\alpha\rightarrow \infty$. \\

\begin{result}{\rm 
\[
\lim_{\alpha\rightarrow \infty} \int_0^\infty H(\alpha,y) \,\phih(y)\,dy = 0
\]
}
\end{result}
\noindent
{\em Proof:} Since
\[ 
\left|\, \int_0^\infty H(\alpha,y) \,\phih(y)\,dy\,\right| \leq  \int_0^\infty H(\alpha,y) \,|\,\phih(y)\,|\,dy,
\]
we show that the right hand side goes to zero as $\alpha$ increases. By the last result, $H(\alpha,y)$ is
a decreasing function of $\alpha$ and therefore 
\[
H(\alpha,y)\, |\,\phih(y)\,| \leq H(1,y)\, |\,\phih(y)\,|
\]
for any $\alpha\geq 1$ and $y\geq 0$. But the right hand side has a finite integral over $[0,\infty)$ because
$\phih(Y)$ is a finite variance estimator.  Hence, by Lebesgue's dominated convergence theorem
\[
\lim_{\alpha\rightarrow \infty}\, \int_0^\infty H(\alpha,y) \,|\,\phih(y)\,|\,dy =
\int_0^\infty \lim_{\alpha\rightarrow \infty} H(\alpha,y) \,|\,\phih(y)\,|\,dy = 0 .
\]
\begin{result}[a]{\rm
If $\phih(y)$ is bounded in a neighborhood of zero, then
\[
\lim_{\alpha\rightarrow \infty} \int_{-\ellp/\alpha}^{0} H(\alpha,y) \,\phih(y)\,dy = 0
\]
}
\end{result}
\noindent
{\em Proof:} First note that for any $y\in (-\ellp/\alpha, 0)$ and $\alpha>1$
\begin{eqnarray*}
\frac{G(\alpha,y)}{G(1,y)} & = & e^{-y(\alpha-1)/(2\sigma^2\phi)}\,\frac
{ e^{\gamma\sqrt{\alpha y + \ellp}} +e^{-\gamma\sqrt{\alpha y + \ellp} }}
{ e^{\gamma\sqrt{y + \ellp}} + e^{-\gamma\sqrt{y + \ellp}} }\\
& \leq & \frac{1}{2} e^{\ellp/(2\sigma^2\phi) + \gamma\sqrt{\ellp}}\,\left(\,e^{2\gamma\sqrt{\ellp}} + 1\,\right)= K_\phi 
< \infty,
\end{eqnarray*}
where $\gamma = 1/(2\sigma^2\phi^{3/2})$. But since $G(1,y)$ is bounded on $(-\ellp/\alpha, 0)$, there
are constants $A_\phi$ and $B_\phi$ such that
\[
0 \leq  A_\phi \leq G(\alpha,y) \leq B_\phi
\]
for all $\alpha>1$ and $y\in (-\ellp/\alpha, 0)$. It follows that
\[
 \int_{-\ellp/\alpha}^{0} H(\alpha,y) \,|\,\phih(y)\,|\,dy \leq 
B_\phi\, \int_{-\ellp/\alpha}^{0} \frac{|\,\phih(y)\,|}{\sqrt{\alpha y + \ellp}}\,dy
\]
Since $\phih(y)$ is bounded in a neighborhood of zero, there is a constant $C>0$ such that for
$\alpha$ large enough
\begin{eqnarray*}
\int_{-\ellp/\alpha}^{0} \frac{|\,\phih(y)\,|}{\sqrt{\alpha y + \ellp}}\,dy & \leq &
C \int_{-\ellp/\alpha}^{0} \frac{dy}{\sqrt{\alpha y + \ellp}}\\
 & = & \frac{C}{\alpha} \, \sqrt{\ellp} \rightarrow 0
\end{eqnarray*}
as $\alpha\rightarrow \infty$. 

\noindent
{\bf Result 3 (b)} If near the origin $\phih(y)$  behaves as $1/|\,y\,|^\beta$ for some
$\beta\in (0,1)$, then
\[
\lim_{\alpha\rightarrow \infty} \int_{-\ellp/\alpha}^{0} H(\alpha,y) \,\phih(y)\,dy = 0
\]
\noindent
{\em Proof:} For $\alpha$ large enough we have the integral
\begin{eqnarray*}
\int_{-\ellp/\alpha}^0 \frac{dy}{|\,y\,|^\beta\,\sqrt{\alpha y + \ellp}} & = & \alpha^{\beta-1} \int_0^\ellp
\frac{dy}{y^\beta\,\sqrt{\ellp - y}}\\
 & = & \alpha^{\beta-1} \left(\,\int_0^{\ellp/2} \frac{dy}{y^\beta\,\sqrt{\ellp - y}} +
\int_{\ellp/2}^\ellp\frac{dy}{y^\beta\,\sqrt{\ellp - y}}\,\right)\\
& \leq & \alpha^{\beta-1}\left(\,\frac{1}{\sqrt{\ellp}}\int_0^{\ellp/2} \frac{dy}{y^\beta} + 
\left(\,\frac{2}{\ellp}\,\right)^\beta\!\! \int_{\ellp/2}^\ellp \frac{dy}{\sqrt{\ellp - y}}\,\right) \\
& = & C\,\alpha^{\beta-1} \rightarrow 0
\end{eqnarray*}
as $\alpha\rightarrow \infty$ for $\beta\in(0,1)$.\\

\noindent
{\bf Remark:} The condition that $\phih$ behaves as $1/|\,y\,|^\beta$ near the origin is not very restrictive
for the following reason. We have just shown that $\Ex (\,1/|\,y\,|^\beta\,)< \infty$ for any $\beta\in (0,1)$
and since $\phih$ has finite variance, it follows from Cauchy-Schwarz inequality that
\[
\Ex \left(\,\frac{|\,\phih(Y)\,|}{|\,Y\,|^\beta}\,\right) \leq \sqrt{\,
\Ex \left(\,\frac{1}{|\,Y\,|^{2\beta}}\,\right)\,\Ex \,\phih(Y)^2\,}.
\]
It follows from this inequality that
\[
\int_{-\ellp/\alpha}^0 \frac{|\,\phih(y)\,|}{|\,y\,|^{\beta/2}\,\sqrt{\alpha y + \ellp}}\, dy < \infty,
\]
which already provides enough information to arrive at the same conclusion of Result 3(b) using
a Mellin transform analysis (e.g., Bleistein \& Handelsman \cite{bleist}).\\

\begin{result}{\rm 
\[
\lim_{\alpha\rightarrow \infty} \Ex \,\phih_\alpha(Y) = 0 \neq \phi,
\]
which contradicts the fact that $\phih_\alpha(Y)$ is an unbiased estimator of $\phi$ for any $\alpha\neq 0$.
}
\end{result}
\noindent
{\em Proof:} We split the integral in \eqref{eq:exphial} into two integrals with
ranges $(-\ellp/\alpha,0)$ and $(0,\infty)$ and then apply Results 2 and 3.

\end{document}